\documentclass{aastex}
\usepackage{amsmath,amssymb,bm}
\begin{document}

\title{The Kozai--Lidov Mechanism in Hydrodynamical Disks. III. Effects of Disk Mass and Self-Gravity}
\author{Wen Fu{$^{1,2}$}, Stephen H. Lubow{$^3$} and Rebecca G. Martin{$^4$}}
\affil{{$^1$}Department of Physics and Astronomy, Rice University, Houston, TX 77005, USA; wf5@rice.edu\\
{$^2$}Los Alamos National Laboratory, Los Alamos, NM 87545, USA\\
{$^3$}Space Telescope Science Institute, Baltimore, MD 21218, USA\\
{$^4$}Department of Physics and Astronomy, University of Nevada, Las Vegas, Las Vegas, NV 89154, USA}


\begin{abstract}
Previously we showed that a substantially misaligned viscous accretion disk with pressure that orbits around one component of a binary system can undergo global damped Kozai-Lidov (KL) oscillations. These oscillations produce periodic exchanges of the disk eccentricity with inclination.  The disk KL mechanism is quite robust and operates over a wide range of binary and disk parameters. However, the effects of self-gravity, which are expected to suppress the KL oscillations for sufficiently massive disks, were ignored.   Here, we analyze the effects of disk self-gravity by means of hydrodynamic simulations and compare the results with the expectations of analytic theory. The disk mass required for suppression in the simulations is a few percent of the mass of the central star and this roughly agrees with an analytical estimate. The  conditions for suppression of the KL oscillations in the simulations are  close to requiring that the disk be gravitationally unstable. We discuss some implications of our results for the dynamics of protoplanetary disks and the related planet formation. 
\end{abstract}


\section{Introduction}


Estimates suggest that 40\% to 50\% of observed exoplanets are in binary systems \citep{Horch2014}. As the birthplace of the planets,  protoplanetary disks are fundamental to explaining planet formation and thus it is important to understand their evolution in a binary system.  Recent ALMA observations revealed large mutual inclinations (greater than $60^{\circ}$) between the circumstellar disks around binary system components \citep[e.g.,][]{Jensen14, Williams14}. Although the binary orbital planes in these systems have not yet been identified, it is possible that at least one of the disks in each system is significantly inclined ( $> 45^\circ$) with respect to the binary orbit. 

The Kozai-Lidov (KL) mechanism is a well known process that occurs when  a ballistic object orbits around one component of a binary  system and the object's orbital plane is substantially misaligned ($i>i_{\rm cr}=39^\circ$) with respect to the binary orbital plane \citep{Kozai62, Lidov62}. In this process, the orbital eccentricity and inclination of the object undergo periodical exchange.  An object on an initially circular
orbit periodically gains and loses eccentricity.
There have been a large number of works on this mechanism since its first discovery in the 1960s \citep[e.g.,][]{Holman97, Innanen97, Kiseleva98, Ford00, Lithwick11, Naoz13a, Liu15, Antognini15} and the KL mechanism has found applications in various astronomical processes, most notably here on the formation of hot Jupiter planets  \citep[e.g.,][]{Wu03, Takeda05, Fabrycky07, Naoz13b, Petrovich15, Rice15}. 

While the KL cycle of orbiting objects has been extensively studied, only very recently was it found to operate on a fluid disk 
with pressure and viscosity by means of hydrodynamic simulations, \citep[][hereafter Paper I]{Martin14}.  Thus, an initially circular disk can become highly eccentric if its initial
tilt exceeds about $45^\circ$ with respect to the binary orbital plane  \citep[hereafter Paper II]{Fu15}. Due to the efficient radial communication via either disk pressure or viscosity, the disk KL cycle for a typical protostellar disk operates in a global, coherent fashion, and the disk remains quite flat (unwarped) throughout the process. The eccentricity growth is fairly uniform in radius across the disk. Unlike the ballistic object case, the disk oscillations are damped due to viscous dissipation, likely involving shocks in the disk.  When the oscillations fully damp, the disk is tilted at the critical KL angle ($i_{\rm cr}\approx 40^\circ$) or somewhat less and the disk is approximately circular. After this stage, the circular disk evolves towards alignment with the binary orbital plane through tidal dissipation effects associated with turbulent viscosity  \citep[e.g.,][]{Kingetal2013}.  Paper II  extended the work of Paper I by surveying a large range of binary and disk parameter space, including the initial disk inclination, temperature, viscosity, size, surface density profile, and the binary mass ratio and eccentricity. The disk KL cycle is a fairly robust process that can occur under many different binary-disk conditions. However, Paper II did not consider the effects of self-gravity.

At early times in the protoplanetary disk evolution, the disk mass is large, several percent of the mass of the central star, and the disk may  
experience dynamical effects of self-gravity. Depending on the properties of the disk, it is possible that self--gravity may be sufficiently strong to prevent KL oscillations from operating. Instead,  the disk will globally precess as a circular object  (as it does for lower inclination disks) \citep{Batygin11,Batygin12,Lai14}. 
This suppression is expected when the disk apsidal precession rate 
due to self--gravity dominates over the apsidal precession rate due to the gravitational effects of the binary that cause the KL oscillations. 
The suppression may not continue as the disk evolves.
As material drains on to the central star, the self-gravity weakens and the KL cycles can begin if the disk remains sufficiently misaligned. 

Differential precession due to  the binary acts to disrupt a disk. 
For a disk to precess coherently, internal torques are required to operate globally and communicate
faster  than the differential precession timescale \citep{Larwoodetal1996}.
The disk model of \cite{Batygin12} maintained disk coherence by means of disk self-gravity and omitted the effects of pressure and viscosity. 
\cite{XG14} pointed out that pressure alone allows a disk to precess coherently and thus self--gravity is not necessarily required. 
In this paper, for the first time we present results on the long term evolution of misaligned self-gravitating disks that have pressure and viscosity.


\section{Numerical Simulations}\label{sec:simulation}

In this section, we describe three-dimensional simulations of an inclined fluid disk around one component of an equal mass binary system, where the binary orbit is circular. We use the smoothed particle hydrodynamic (SPH) code \texttt{PHANTOM} \citep{Lodato10, Price10, Nixon12a, Nixon12b, Price12, Nixon13}. The simulation setup is similar to that in Paper II. An initially circular disk orbits around a central binary component of mass $M_{\rm c}$ under the influence of the perturber binary component of mass $M_{\rm p}$.  The disk and stars interact gravitationally and respond to these interactions. 
In particular, the orbit
of the binary is affected by its gravitational interactions with the disk, although the changes to the binary orbit are small. 
The orbital plane of the disk is initially inclined to the binary orbital plane by $50^{\circ}$. We adopt a locally isothermal equation of state and an explicit accretion disk viscosity. We include a nonlinear term with a coefficient $\beta_{\rm AV}=2$ (AV stands for artificial viscosity) in order to suppress interparticle penetration, as is standard in SPH codes.   The disk sound speed  is $c_{\rm s} \propto r^{-3/4}$ and the initial surface density distribution is $\Sigma \propto r^{-3/2}$, such that both $\alpha$ \citep{SS73} and the smoothing length $\left<h\right> /H$  are constant over the disk radius, $r$ \citep{Lodato07}. We start the simulations with $1 \times 10^6$ SPH particles in the disk.
The \texttt{PHANTOM} code adopts a cubic spline kernel as the smoothing kernel. The number of neighbors is roughly constant at $N_{\rm neigh} \approx 58$. The disk initially extends from radius $r_{\mathrm{in}}$ to radius $r_{\mathrm{out}}$. The inner boundary of the simulated region is set to the initial disk inner disk radius $r_{\mathrm{in}}$. As particles move to $r \le r_{\mathrm{in}}$, they are removed from the simulation while their mass and momentum are added to the central star. We also impose an inner boundary radius around the perturbing companion, since some disk mass can be transferred to that component. In terms of the binary orbital size $a_{\rm b}$, the initial disk inner and outer radii in our simulations are located at $r_{\rm in}=0.025 a_{\rm b}$ and $r_{\rm out}=0.25 a_{\rm b}$, respectively. The initial outer disk radius is chosen to be the tidal truncation radius of a coplanar disk \citep{Paczynski77}. However, because the tidal torques on a misaligned disk are weaker \citep{Lubow15}, the disk initially expands somewhat  beyond this radius.

We describe simulations with two sets of parameters for the disk aspect ratio and viscosity. The first is $H/R=0.1$ and $\alpha=0.01$ and the second is $H/R=0.06$ and $\alpha=0.01$ ($H/R$ is evaluated at disk inner edge $r_{\rm in}$). The disk is resolved with shell-averaged smoothing length per scale height $\left<h\right>/H\approx 0.26$ and $0.37$, respectively. For each set of $H/R$ and $\alpha$, we consider three different initial disk masses $M_{\rm d}=0.001M$, $0.01M$ and $0.03M$ (in units of the total binary mass $M=M_{\rm c}+M_{\rm p}$). In Papers I and II, we ignored disk self-gravity given the low disk mass used there ($0.001M$). In this paper, we take into account the effect of disk self-gravity and compare simulations with and without self-gravity for the same disk mass. The algorithm for the SPH implementation of self-gravity in \texttt{PHANTOM} is described in \cite{Price07b}, which discusses numerical tests based on the radial oscillations and the static structure of a polytrope. The self-gravity algorithm of \texttt{PHANTOM} has also been used to study the formation of giant molecular clouds \citep{Dobbs08} and star clusters \citep{Price08}.


\begin{deluxetable}{l c r}
\tabletypesize{\footnotesize}
\tablecolumns{4}
\tablewidth{0pt}
\tablecaption{Parameters of the SPH simulations for equal mass binary systems with total mass of $M$ and separation of $a_{\rm b}$\label{table:parameters}}
\tablehead{
\colhead{Binary and Disk parameters} & 
\colhead{Symbol} & 
\colhead{Value} 
}
\startdata
Mass of each binary component & $M_{\rm p}/M=M_{\rm c}/M$ & 0.5\\
Binary orbital eccentricity & $e_{\rm b}$ & 0\\
Initial number of particles & $N$ & $10^6$\\
Initial disk mass & $M_{\mathrm{disk}}/M$  & [0.001, 0.01, 0.03] \\
Initial disk outer radius & $r_{\mathrm{out}}/a_{\rm b}$ & 0.25\\
Initial disk inner radius & $r_{\mathrm{in}}/a_{\rm b}$ & 0.025\\
Mass accretion radius & $r_{\mathrm{acc}}/a_{\rm b}$ & 0.025\\
Disk viscosity parameter & $\alpha$ & [0.01, 0.1]\\
Disk aspect ratio & $H/r\, (r=r_{\mathrm{in}})$ & [0.06, 0.1]\\
Initial disk surface density profile $\Sigma \propto r^{-\gamma}$ & $\gamma$ & 1.5\\
Initial disk inclination & $i_{\rm 0}$ & $\mathrm{50^{\circ}}$\\
\enddata
\label{tab:params}
\end{deluxetable}

\begin{figure}
\centering
\includegraphics[width=0.7\textwidth]{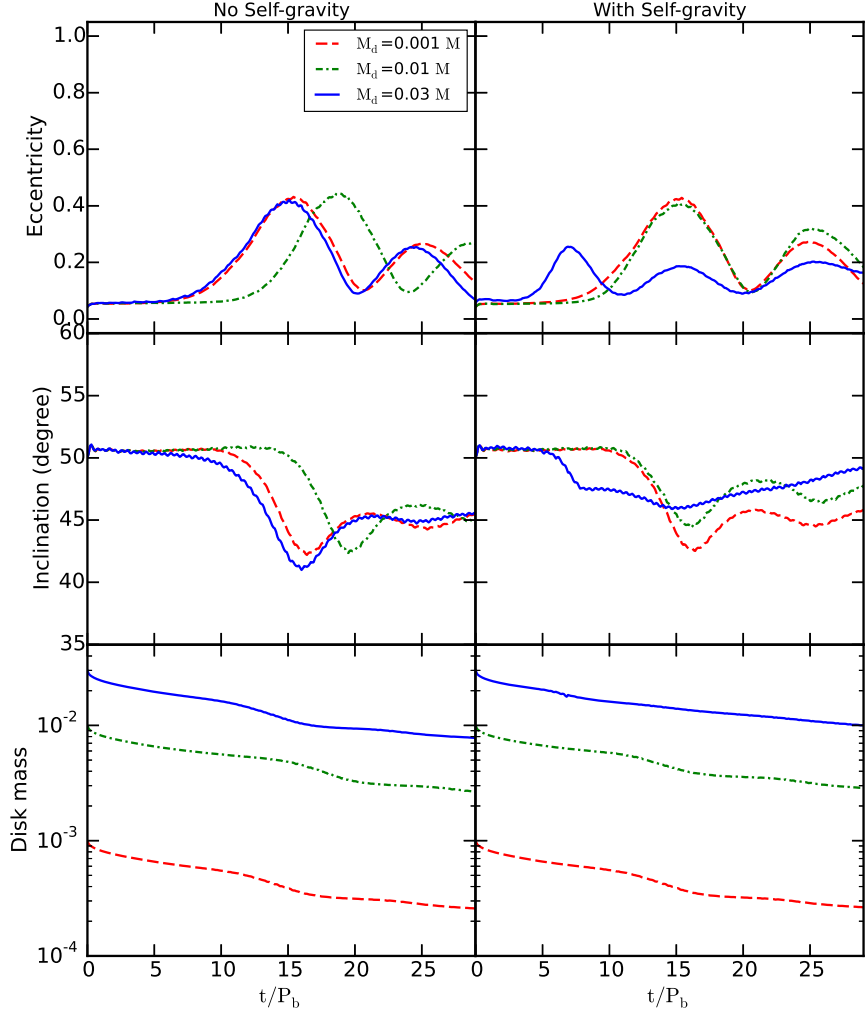}
\caption{Evolution of the eccentricity (upper row), inclination (middle row), and total mass (bottom row) of the circumprimary disk. The left column is for simulations without disk self-gravity, whereas the right column is for simulations that take into account disk self-gravity. The eccentricity and inclination are measured at a radius of $r=0.2a_{\rm b}$ (the initial disk radial range is $0.025a_{\rm b} \leq r \leq 0.25a_{\rm b}$), where $a_{\rm b}$ is the binary semi--major axis and $P_{\rm b}$ is the binary orbital period. Within each plot, each line represents a simulation with a different initial disk mass.  The units of mass are that  of the total binary mass, $M$. All lines are averaged over one binary orbital period.  The initial number of SPH particles is $10^6$. The disk aspect ratio at the inner edge is $H/R=0.1$. The disk viscosity parameter is $\alpha=0.01$ and the initial disk inclination is $i=50^{\circ}$.  \label{fig:fig1}}
\end{figure}

Figure \ref{fig:fig1} shows the evolution of the disk eccentricity (upper row), inclination (middle row), and mass (bottom row) for three different initial disk masses. The eccentricity and inclination are measured at a representative radius $r=0.2a_{\rm b}$, since the disk remains flat and the eccentricity is roughly constant with radius.  The left and right columns show simulations without and with disk self-gravity, respectively. 
In all cases, disk gravity acts on the stars and slightly affects the binary orbit. 
In post-processing the simulation outputs, we compute the orbital eccentricity and inclination for each particle using its position and velocity information, then divide the disk into $100$ radial bins and calculate the mean properties of the particles within each bin. Results for the eccentricity and inclination are taken from the radial bin centered on radius $0.2 a_{\rm b}$. This is similar to the method used in Papers I and II, except now we average the properties over one binary orbit to smooth out the fluctuations on the binary orbital timescale. Note that the initial disk eccentricity is non-zero ($e\simeq 0.07$) in the upper row, even though the disk is initially set up to be circular. This non-zero eccentricity value is a consequence of the way
 we calculate the particle eccentricity and inclination (Equations [7] and [8] in Paper II) that treats particles as ballistic, i.e., assuming particles only feel the gravitational force of the central object, whereas particles actually are also influenced by the disk pressure force. Therefore, the non-zero initial disk eccentricities we see here are purely an artifact of our ignoring the disk pressure in the orbital elements calculation. The effect is more prominent here than in the simulations shown in Papers I and II because here the disk has higher temperature, $H/R=0.1$ (compared with $H/R=0.035$). However, it is still small enough not to affect our general interpretation, since we are focused on the major changes of the disk properties during KL oscillations. 

As seen in the left column of Figure \ref{fig:fig1}, where disk self-gravity is not included, increasing the disk mass does not  significantly affect the disk orbital elements during the KL cycle. The period and amplitude of the oscillations are similar, although there is some variation in the time of the first peak. 
The difference in the onset of the initial KL cycle is due to the fact that disk mass slightly affects the companion's orbit. The timing of the onset is sensitive to the initial conditions. 
Even with $M_{\rm d}=0.03\,M$,  the peak eccentricity value of $e\simeq 0.42$ is almost the same as that for the simulations with $M_{\rm d}=0.001\,M$ and $0.01\,M$ with a peak $e=0.44$. As seen in the right column of the figure, for the lower mass disks, self-gravity generally has little effect on disk eccentricity and inclination. However, the effects of the KL cycle are significantly reduced at the highest disk mass of $M_{\rm d}=0.03M$ , which has an initial disk minimum Toomre Q $\sim 2.6$. The peak eccentricity in that case is $e\simeq 0.2$ compared to $e\simeq 0.42$ with $M_{\rm d}=0.001M$ and $0.01M$.  The bottom row shows that by the end of the simulation at time $t=29 P_{\rm b}$, the disk in general has lost more than half of its initial mass. The majority of the lost mass is accreted on to the central object. A small fraction of the mass is ejected from the disk, mostly ending up around the companion binary star. Comparing simulations with the same initial disk mass, the run with self-gravity has more mass remaining at the end of the simulation than the one without self-gravity. This effect may be a consequence of the disk self-gravity helping to retain particles, and also reducing the particle loss at inner disk boundary, since the KL cycle is weakened. 

\begin{figure}
\centering
\includegraphics[width=0.95\textwidth]{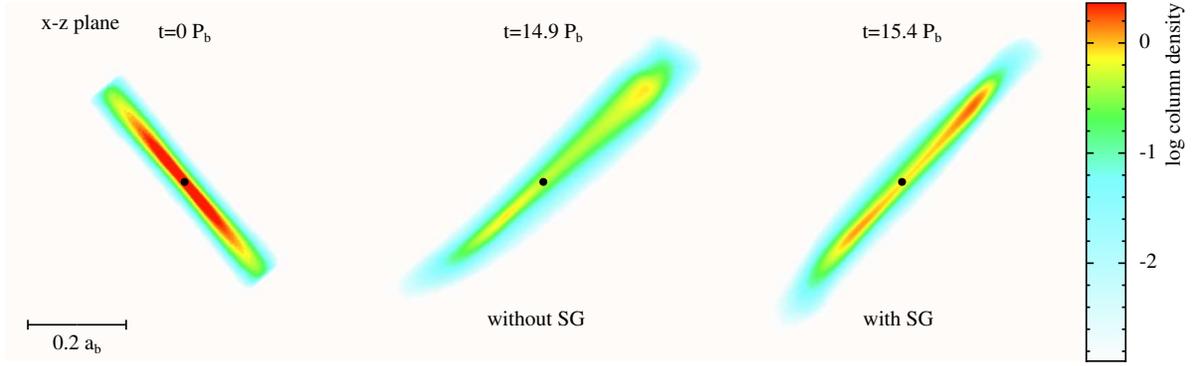}
\caption{View of the disk towards the $x-z$ plane for the two simulations with $M_{\rm d}=0.03\,M$, $H/R=0.1$ and $\alpha=0.01$, with and without self--gravity. The binary orbit is in the $x-y$ plane (i.e., the perturbing object moves into and out of the page).  The left panel shows the initial disk setup that is the same for both simulations. The middle (right) panel shows the disk without (with)  self-gravity after it has undergone nodal precession by $180^\circ$.  The central mass is denoted by the black dot. The color coding is for the logarithm base $10$ of the column density (i.e., density integrated along the line of sight) in units of $(M_{\rm c}+M_{\rm p})/a^2_{\rm b}$.\label{fig:fig2} (Color online)}
\end{figure}

Figure \ref{fig:fig2} shows edge-on views of the disk in the two simulations with $M_{\rm d}=0.03\,M$, one with and the other without disk self-gravity. The left panel shows the initial conditions that were applied to both simulations.
The middle and right panels are edge-on views after the disks have undergone nodal precession by $180^\circ$.
The middle panel is from the run without disk self-gravity  (the blue curves in the left column of Figure \ref{fig:fig1}) at a time of $t=14.9 P_{\rm b}$, while the right panel is from the run that includes disk self-gravity (the blue curves in the right column of Figure \ref{fig:fig1}) at a time of $t=15.4P_{\rm b}$. These runs have different disk nodal precession rates and as a result the middle and right panels are shown at slightly different epochs. Without self-gravity, the middle panel shows a substantially lopsided disk due to the KL-driven eccentricity growth. 
In the right panel, with disk self-gravity, the disk is still eccentric, but not as eccentric as in the middle panel. 
Although at an earlier time ($t\simeq 7 P_{\rm b}$) the disk has a slightly higher eccentricity, the average disk eccentricity is clearly lower when self-gravity is included (see comparison of the blue curves in the top two panels of Figure \ref{fig:fig1}). The main point here is that
in both panels, the disk remains fairly flat (no significant warping) due to the efficient radial communication (provided mainly by pressure in these cases; see discussions in Paper II). 

\begin{figure}
\centering
\includegraphics[width=0.8\textwidth]{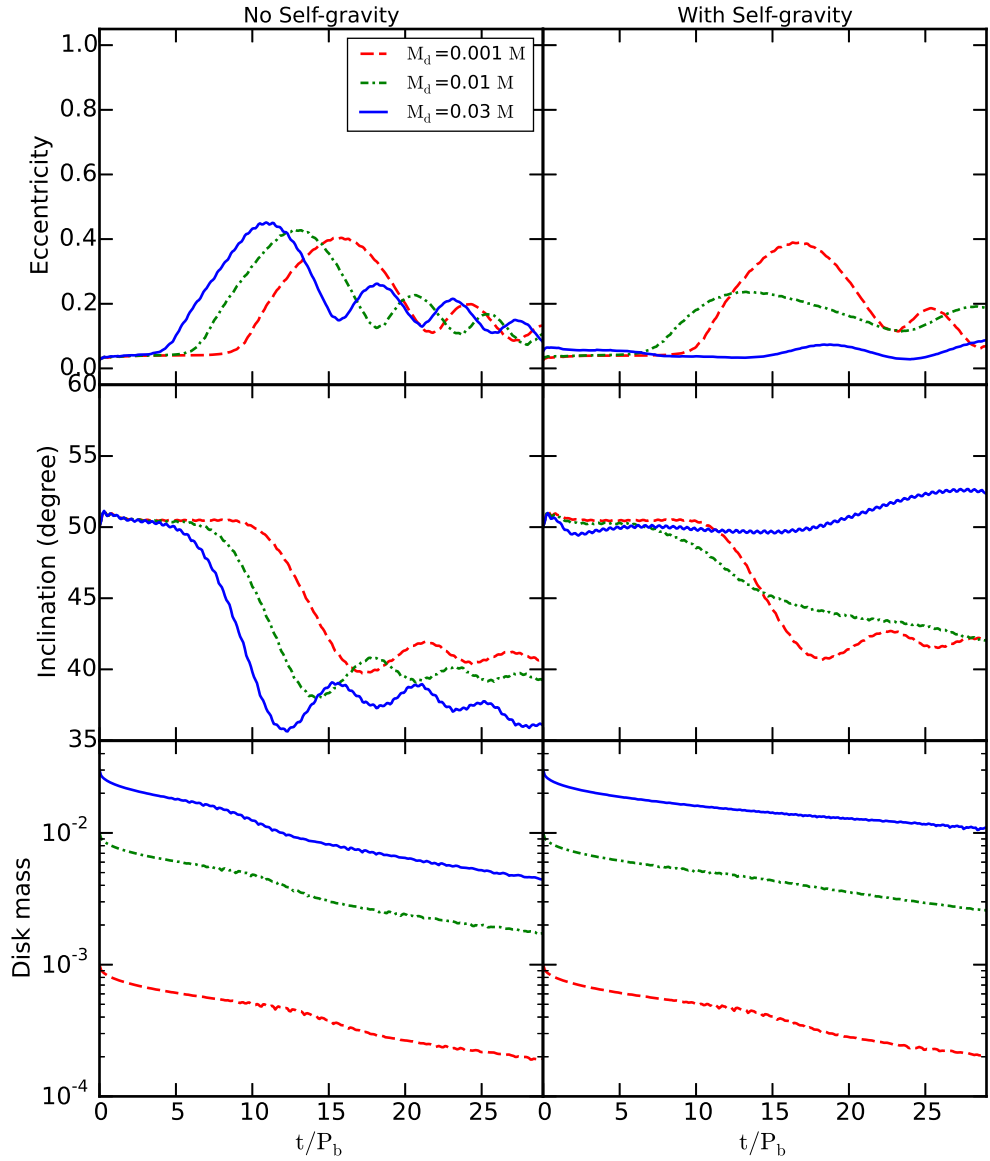}
\caption{Similar to Figure \ref{fig:fig1} except for different disk aspect ratio $H/R=0.06$ and viscosity parameter $\alpha=0.1$. \label{fig:fig3}}
\end{figure}

Figure \ref{fig:fig3} is similar to Figure \ref{fig:fig1} except for a different disk aspect ratio, $H/R=0.06$, and viscosity, $\alpha=0.1$. When disk self-gravity is not included, we see again that the disk mass has only a minor effect on the disk KL cycle (left column). 
The oscillation timescales and amplitudes are similar.
As we discussed in Figure \ref{fig:fig1}, the reason why the disk mass plays a role in the absence of self-gravity is that the disk mass slightly affects the orbit of the companion, which in turn affects the onset of the disk KL cycle. With self-gravity included, we see a greater suppression of the disk KL cycle by self-gravity than that in Figure \ref{fig:fig1}. With $M_{\rm d}=0.03\,M$, the KL mechanism barely operates. For the disk mass of $M_{\rm d}=0.01\,M$, there is a significant reduction. The lowest mass disk remains relatively unaffected by self-gravity. Note that for the highest disk mass, the initial disk has a minimum Toomre Q $\simeq 1.6$, which puts the disk on the verge of becoming gravitationally unstable to non-axisymmetric disturbances. In this paper, we have restricted our attention to simulations in which we do not see any sign of gravitational instability, such as fragmentation or clump formation.

\section{Discussion and Conclusions}\label{sec:summary}

Self-gravity introduces a source of  disk apsidal precession, in addition to that due to the binary companion that causes the KL oscillations. If the self-gravity induced apsidal disk precession rate is faster than the binary induced precession rate, then the  disk KL cycle can be suppressed \citep[e.g.,][]{Holman97}. We estimate the magnitude of the local self-gravity induced disk apsidal precession rate $|g_{\rm sg}(r)| \simeq \pi G \Sigma(r)/ (\Omega(r) r)$  from Equation (1) of \cite{Tremaine98}, which is consistent with the rate  given in \cite{Batygin11}.  We determine the binary companion induced disk  local precession rate $g_{\rm b}(r)$ by approximating the companion's potential as arising from a uniform ring of mass $M_{\rm p}$ and radius $a_{\rm b}$, where $a_b$ is binary separation. Since the disk maintains its coherence, its global precession rate is crudely estimated as the angular-momentum-weighted average of the local rates (see Equation (9) of Paper II).
For the equal mass binary system we consider, we take $r_{\rm in}=0.025 a_b$, $r_{\rm out}=0.3 a_b$ (due to disk expansion from the initial $r_{\rm out}$). The global disk precession rate caused by a binary companion is $\overline{g}_{\rm b}\sim 0.03 \Omega_{\rm b}$, where $\Omega_{\rm b}$ is the binary orbital frequency. This precession rate is similar in value to the nodal precession rate implied by Figure  \ref{fig:fig2} that shows that the disks undergo half the nodal precession  period at a time of about $15 P_{\rm b}$. The global disk precession rate due to self-gravity is $|\overline{g}_{\rm sg}| \sim  0.008(M_{\rm d}/0.001 M) \Omega_{\rm b}$. Requiring $|\overline{g}_{\rm sg}| \geq |\overline{g}_{\rm b}|$ provides an estimate of the minimum disk mass to suppress KL oscillations as $M_{\rm d} \sim 0.004 M$.  According to Figure 7 of \cite{Batygin11}, the critical disk mass for suppressing KL  oscillations is generally somewhat larger than this value, depending on the disk
inclination. Based on that work, for the initial disk inclination we have used, the estimated disk mass for KL suppression is about a
factor of 5 times larger,  that is  $M_{\rm d} \sim  0.02M$.

In the case of $H/R=0.1$, the blue curve in the right column of Figure \ref{fig:fig1} clearly shows that the disk KL cycle is weakened for the case of an initial disk mass of $0.03\,M$. In spite of the reduced oscillation amplitude, the disk still undergoes KL cycles. There is an eccentricity peak and inclination valley at  a time $t \sim 7P_{\rm b}$ when the disk has a mass of about $M_{\rm d}\simeq 0.02M$ which we adopt as a crude estimate for the minimum mass for KL suppression.   

In the case of $H/R=0.06$, the blue curve in the right column of Figure \ref{fig:fig3}  shows that the KL cycle is suppressed, for the case of an initial disk mass of $0.03 M$. In the case of the lower initial disk mass of $0.01M$,
the KL oscillation appears to be present, but weakened.
A reasonable estimate of the critical disk mass suppression of KL oscillations is  similar to our estimate for the other disk model, $\sim 0.02M$  
. 
Therefore, the simulations show that the critical disk mass for suppressing KL oscillations 
is in the range of $M_{\rm d}\sim 0.02M$ ( 4\% of the mass of the central star) that is about a factor of 5 larger than the value based
on equating the self-gravity induced precession rate with the binary induced precession rate.
It is in agreement with the value implied by  \cite{Batygin11} that is discussed above.

Papers I and II showed that global disk KL oscillations damp. After damping,  the disk inclination  is at or below critical KL angle. 
If a planet forms after this stage, then its initial orbital tilt is not large enough to trigger the KL mechanism. 
This reduced tilt poses a challenge to the KL model of planets which assumes initially highly inclined planet orbits. 
 The tilt evolution of such planet--disk configurations requires further study.

Suppression of the KL cycle by means of self-gravity requires substantial disk masses. At the critical disk mass we found, the disk is quite close to being gravitationally unstable. In other words, we find a relatively narrow window of disk masses where the disk KL oscillation is significantly subdued, while the disk is still stable to gravitational instability. 

The highest initial disk mass we consider in this paper, $M_{\rm d}=0.03\,M$,  shows more warping with self--gravity included than in cases without self--gravity. Initially, the inclination of the inner parts of the disk increases, while that of the outer parts decreases.  
In Paper II, we found that the critical (minimum) tilt angle for KL oscillations to operate in a disk is about $45^\circ$, somewhat higher than the
critical angle for free particles ($39^\circ$). 
In this paper, we have limited our analysis to an initial disk tilt of $50^\circ$
that is just $5^\circ$ above the minimum value for KL oscillations. Such initial conditions produce much more moderate KL effects
on a disk than occurs at higher tilt angles.  We also reported in Paper II that strong density enhancements are found
during KL oscillations of a non-self-gravitating disk with a larger initial tilt. These density enhancements appear to involve shocks.
With larger initial tilts than we have considered here, a self-gravitating disk that undergoes KL oscillations
may undergo disk fragmentation. Fragmentation of a disk is a possible mechanism for planet formation and may be aided by binary perturbations \citep{Boss1997, Boss06}. 
The disk KL cycle may then provide an alternative means by which a
binary companion can promote disk fragmentation/clumping.  Such effects should be explored in the future.


W.F. and S.H.L. acknowledge support from NASA grant NNX11AK61G.  Computing resources supporting this work were provided by the institutional computing program at Los Alamos National Laboratory. We thank Daniel Price for providing the \texttt{PHANTOM} code for SPH simulations and \texttt{SPLASH} code \citep{Price07a} for data analysis and rendering of figures.



\begin{thebibliography}{}
\bibitem[Antognini(2015)]{Antognini15}
Antognini, J. M. O. 2015, arXiv:1504.05957

\bibitem[Batygin et al.(2011)]{Batygin11}
Batygin, K., Morbidelli, A., \& Tsiganis, K. 2011, A\&A, 533, A7

\bibitem[Batygin(2012)]{Batygin12}
Batygin, K. 2012, \nat, 491, 418

\bibitem[Boss (1997)]{Boss1997}
Boss, A. P. 1997, Science, 276, 1836

\bibitem[Boss (2006)]{Boss06}
Boss, A. P. 2006, \apj, 641, 1148

\bibitem[Dobbs (2008)]{Dobbs08}
Dobbs, C. L. 2008, \mnras, 391, 844

\bibitem[Fabrycky \& Tremaine(2007)]{Fabrycky07}
Fabrycky, D., \& Tremaine, S. 2007, \apj, 669, 1298

\bibitem[Ford et al.(2000)]{Ford00}
Ford, E. B., Kozinsky, B., \& Rasio, F. A. 2000, \apj, 535, 385


\bibitem[Fu et al.(2015)]{Fu15}
Fu, W., Lubow, S. H., \& Martin, R. G. 2015, \apj, 807, 75

\bibitem[Horch et al.(2014)]{Horch2014}
Horch, E. P., Howell, S. B., Everett M. E., Ciardi D. R., 2014, ApJ, 795, 60

\bibitem[Holman et al.(1997)]{Holman97}
Holman, M. J., Touma, J., \& Tremaine, S. 1997, \nat, 386, 254

\bibitem[Innanen et al.(1997)]{Innanen97}
Innanen, K. A., Zheng, J. Q., Mikkola, S., \& Valtonen, M. J. 1997, \aj, 113, 1915

\bibitem[Jensen \& Akeson(2014)]{Jensen14}
Jensen, E. L. N. \& Akeson, R. 2014, \nat, 511, 567

\bibitem[King et al.(2013)]{Kingetal2013}
King, A. R., Livio, M., Lubow, S. H., \& Pringle, J. E. 2013, MNRAS, 431, 2655

\bibitem[Kiseleva et al.(1998)]{Kiseleva98}
Kiseleva, L. G., Eggleton, P. P., \& Mikkola, S. 1998, \mnras, 300, 292
 

\bibitem[Kozai(1962)]{Kozai62}
Kozai, Y. 1962, \aj, 67, 591

\bibitem[{{Larwood} {et~al.}(1996){Larwood}, {Nelson}, {Papaloizou}, \&
  {Terquem}}]{Larwoodetal1996}
{Larwood}, J.~D., {Nelson}, R.~P., {Papaloizou}, J.~C.~B., \& {Terquem}, C.
  1996, MNRAS, 282, 597
  
\bibitem[Lai(2014)]{Lai14}
Lai, D. 2014, MNRAS, 440, 3532

\bibitem[Lidov(1962)]{Lidov62}
Lidov, M. L. 1962, P\&SS, 9, 719

\bibitem[Lithwick \& Naoz(2011)]{Lithwick11}
Lithwick, Y., \& Naoz, S. 2011, \apj, 742, 94

\bibitem[Liu et al.(2015)]{Liu15}
Liu, B., Mu\~{n}oz, D., \& Lai, D. 2015, \mnras, 447, 747

\bibitem[Lodato \& Pringle(2007)]{Lodato07}
Lodato, G., \& Pringle, J. E. 2007, MNRAS, 381, 1287

\bibitem[Lodato \& Price(2010)]{Lodato10}
Lodato, G., \& Price, D. J. 2010, \mnras, 405, 1212


\bibitem[Lubow, Martin \& Nixon(2015)]{Lubow15}
Lubow S. H., Martin R. G., \& Nixon C. J., 2015, \apj, 800, 96


\bibitem[Martin et al.(2014)]{Martin14}
Martin, R. G., Nixon, C., Lubow, S. H., et al. 2014, ApJL, 792, L33




\bibitem[Naoz et al.(2013a)]{Naoz13a}
Naoz, S., Farr, W. M., Lithwick, Y., Rasio, F. A., \& Teyssandier, J. 2013a, \mnras, 431, 2155

\bibitem[Naoz et al.(2013b)]{Naoz13b}
Naoz, S., Kocsis, B., Leob A., \& Yunes, N. 2013b, \apj,773, 187

\bibitem[Nixon et al.(2013)]{Nixon13}
Nixon, C., King, A., \& Price, D. 2013, MNRAS, 434, 1946

\bibitem[Nixon (2012)]{Nixon12a}
Nixon, C. J. 2012, MNRAS, 423, 2597

\bibitem[Nixon \& King (2012)]{Nixon12b}
Nixon, C. J., \& King, A. R. 2012, MNRAS, 421, 1201

\bibitem[Paczynski(1977)]{Paczynski77}
Paczynski B., 1977, ApJ, 216, 822

\bibitem[Petrovich(2015)]{Petrovich15}
Petrovich, C. 2015, \apj, 799, 27

\bibitem[Price(2007)]{Price07a}
Price, D. J. 2007, \pasa, 24, 159

\bibitem[Price \& Bate (2008)]{Price08}
Price, D. J., \& Bate, M. R. 2008, \mnras, 385, 1820

\bibitem[Price \& Monaghan(2007)]{Price07b}
Price, D. J., \& Monaghan, J. J. 2007, \mnras, 374, 1374

\bibitem[Price \& Federrath(2010)]{Price10}
Price, D. J., \& Federrath, C. 2010, \mnras, 406, 1659

\bibitem[Price(2012)]{Price12}
Price, D. J. 2012, JCoPh, 231, 759




\bibitem[Rice (2015)]{Rice15}
Rice, K. 2015, \mnras, 448, 1729

\bibitem[Shakura \& Sunyaev(1973)]{SS73}
Shakura, N. I., \& Sunyaev, R. A., 1973, A\&A, 24, 337


\bibitem[Takeda \& Rasio(2005)]{Takeda05}
Takeda, G., \& Rasio, F. A. 2005, \apj, 627, 1001

\bibitem[Tremaine (1998)]{Tremaine98}
Tremaine, S. 1998, \apj, 116, 2015

\bibitem[Williams et al.(2014)]{Williams14}
Williams, J. P., Mann, R. K., Di Francesco, J., et al. 2014, \apj, 796, 120

\bibitem[Wu \& Murray(2003)]{Wu03}
Wu, Y., \& Murray, N. 2003, \apj, 589, 605

\bibitem[Xiang-Gruess \& Papaloizou(2014)]{XG14}
Xiang-Gruess, M. \& Papaloizou, J. C. B. 2014, MNRAS, 440, 1179



\end{thebibliography}
\end{document}